\documentclass[
aps, prl,
amsmath, amssymb, superscriptaddress,
reprint,
]{revtex4-2}

\usepackage{}
\usepackage{graphicx}
\usepackage{dcolumn}
\usepackage{bm}
\usepackage[mathlines]{lineno}

\usepackage[utf8]{inputenc}
\usepackage[T1]{fontenc}
\usepackage{mathptmx}

\usepackage[separate-uncertainty=true, range-units=repeat, range-phrase={\ to }, list-final-separator = {\ and \ }]{siunitx}
\DeclareSIUnit\fluence{\milli\joule\per\centi\meter\squared}

\usepackage[version=4]{mhchem}
\usepackage{graphicx, float}
\usepackage[capitalise]{cleveref}
\usepackage{tikz}
\usepackage{subfig}

\usepackage{ulem}
\usepackage{color}

\begin{document}
\title{Robust scenario for the generation of non-equilibrium topological fluctuation states}
\author{Kathinka Gerlinger} 
 \thanks{These two authors contributed equally}
 \affiliation{Max Born Institute for Nonlinear Optics and Short Pulse Spectroscopy, 12489 Berlin, Germany}
\author{Rein Liefferink} 
 \thanks{These two authors contributed equally}
 \affiliation{Radboud University, Institute for Molecules and Materials, 6525 AJ Nijmegen, the Netherlands}
\author{Michael Schneider} 
 \affiliation{Max Born Institute for Nonlinear Optics and Short Pulse Spectroscopy, 12489 Berlin, Germany}
\author{Lisa-Marie Kern} 
 \affiliation{Max Born Institute for Nonlinear Optics and Short Pulse Spectroscopy, 12489 Berlin, Germany}
\author{Christopher Klose} 
 \affiliation{Max Born Institute for Nonlinear Optics and Short Pulse Spectroscopy, 12489 Berlin, Germany}
\author{Daniel Metternich} 
 \affiliation{Helmholtz-Zentrum Berlin, 14109 Berlin, Germany}
 
\author{Dieter Engel} 
 \affiliation{Max Born Institute for Nonlinear Optics and Short Pulse Spectroscopy, 12489 Berlin, Germany}

\author{Flavio Capotondi} 
 \affiliation{Elettra Sincrotrone Trieste, 34149 Basovizza TS, Italy}
\author{Dario De Angelis} 
 \affiliation{Elettra Sincrotrone Trieste, 34149 Basovizza TS, Italy}
\author{Matteo Pancaldi} 
 \affiliation{Elettra Sincrotrone Trieste, 34149 Basovizza TS, Italy}
\author{Emanuele Pedersoli} 
 \affiliation{Elettra Sincrotrone Trieste, 34149 Basovizza TS, Italy}

\author{Felix Büttner} 
 \affiliation{Helmholtz-Zentrum Berlin, 14109 Berlin, Germany}
\author{Stefan Eisebitt} 
 \affiliation{Max Born Institute for Nonlinear Optics and Short Pulse Spectroscopy, 12489 Berlin, Germany}
 \affiliation{Technische Universität Berlin, Institut für Optik und Atomare Physik, 10623 Berlin, Germany}
\author{Johan H. Mentink} 
 \email{j.mentink@science.ru.nl}
 \affiliation{Radboud University, Institute for Molecules and Materials, 6525 AJ Nijmegen, the Netherlands}
\author{Bastian Pfau} 
 \email{bastian.pfau@mbi-berlin.de}
 \affiliation{Max Born Institute for Nonlinear Optics and Short Pulse Spectroscopy, 12489 Berlin, Germany}
 
\date{\today}

\begin{abstract}
The recently discovered topological fluctuation state provides a fascinating new perspective on the ultrafast emergence of topology in condensed matter systems.
However, rather little is known about the physics of this state and the origin of the topological fluctuations.
Using time-resolved small-angle x-ray scattering, we observe that topological fluctuation states appear after laser excitation even if the final state does not host stable skyrmions.
Simulations support these findings and reveal that the fluctuations originate from the competition between spontaneous nucleation and decay of skyrmions, consistent with Arrhenius-like activation over a potential barrier.
Stable skyrmions can freeze out of such fluctuations when the effective temperature of the system relaxes faster than the decay time of the skyrmions.
Our results reveal a robust scenario for the generation of topological fluctuation states, potentially enabling their study in a wide variety of magnetic systems.
\end{abstract}

\maketitle


The emergence of nanometer-scale magnetic textures in thin-film magnetic multilayers with perpendicular anisotropy has fascinated scientists for a long time and stimulated widespread fundamental and applied research on these customizable materials \cite{Hellwig2007, Barla2021}.
This research has even intensified in the last years by the discovery of topological textures, so-called skyrmions, which stabilize in the material, even at room temperature, by competing interactions \cite{Moreau-Luchaire2016, Boulle2016, Woo2016, Wiesendanger2016}.
Typically, however, the skyrmion state remains hidden during adiabatic field cycling \cite{Woo2016} due to the large topological energy barrier associated with its nucleation.

The most common way to create skyrmions is the application of spin torques generated from spin-polarized currents \cite{Romming2013, Woo2016, Buttner2017, Finizio2019}.
Recently, though, it was demonstrated that also ultrashort laser excitation leads to the formation of a skyrmion state in the material \cite{Je2018, Buttner2020} if the laser fluence exceeds a certain nucleation threshold \cite{Buttner2020, Gerlinger2021}.
This topological phase transition proceeds on a timescale of only 300 ps, despite the massive reorientation of spins required to nucleated areas of switched magnetization in the ferromagnet \cite{Buttner2020}. 
Even more important, this research provided the first evidence of the existence of a so-called fluctuation state which mediates the topological switching. 
Previous equilibrium Monte-Carlo simulations \cite{Rozsa2016, Boettcher2018}, have elucidated similar fluctuating states in the equilibrium phase diagram, localized between field-polarized ferromagnetic and paramagnetic phases.
Still, very little is known about the physics of topological fluctuations states out of equilibrium, particularly their role in topological phase transitions and the fundamental speed limit of such transitions. 
For example, previous work explained the topological switching in the fluctuation state mainly by an effective elimination of the topological energy barrier \cite{Buttner2020}. 
However, it remains unclear how the symmetry breaking towards stable skyrmions proceeds dynamically and if this process necessitates non-equilibrium conditions.

Here, we report on details of the fluctuation state accessible in Co/Pt multilayers after laser excitation. 
Combining resonant scattering experiments and atomistic spin simulations on ultrafast timescales, we find that the system robustly passes through the topological fluctuation state---even when the topology of the final state is trivial. 
We identify the fluctuation state as a competition between nucleation and decay of skyrmions. 
Its dynamics can be explained with Arrhenius-like activation over an energy barrier over a broad range of temperatures. 
Based on field and temperature-dependent relaxation times, we model the topological fluctuations with a basic rate equation, even under highly non-equilibrium conditions and on timescales much shorter than typical micromagnetic timescales. 
As a result, stable skyrmions form when the effective temperature of the system relaxes faster than the decay time of the skyrmions, opening up new possibilities to approach the speed limit for the topological phase transition.


\begin{figure}
    \includegraphics[width=\columnwidth]{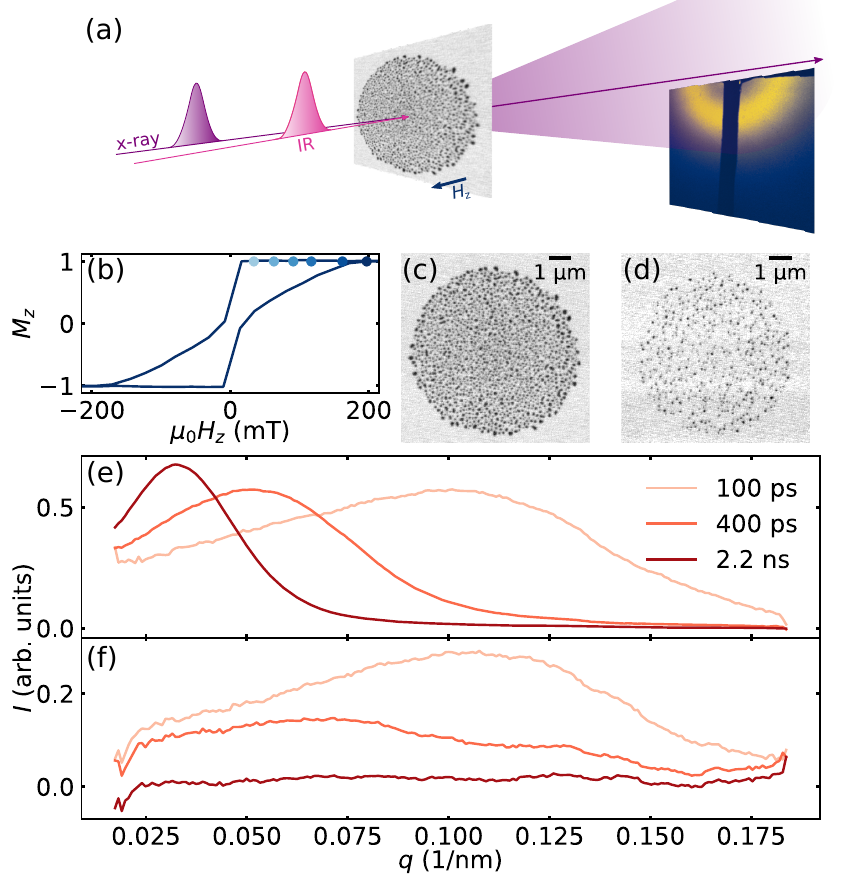}
    \caption{Detection of the topological magnetic states. (a) Illustration of the pump--probe setup.
    (b) The out-of-plane magnetic hysteresis of the multilayer film. The blue circles indicate the fields at which time-resolved data were acquired.
    (c,d) X-ray microscopy images of the skyrmion state after exposure with one IR pulse at an applied field of \SI{30}{\milli\tesla} and \SI{60}{\milli\tesla}, respectively.
    (e,f) Scattered intensity as function of momentum transfer $q$ at delay times indicated and under an applied field of \SI{34}{\milli\tesla} and \SI{197}{\milli\tesla}, respectively.
    For better comparability, the spectra at \SI{400}{\pico\second} and \SI{2.2}{\nano\second} in (e) were divided by 4 and 10, respectively.}
    \label{fig:expdata}
\end{figure}

Our magnetic multilayer sample consists of \num{12} repeats of Co(\SI{0.6}{nm})/Pt(\SI{0.8}{nm}) bilayers and exhibits a typical sheared hysteresis of a thin film ferromagnet with perpendicular anisotropy, forming lateral magnetic domains stabilized by stray fields (Fig.~\ref{fig:expdata}(b), see supplemental material (SM), section I).
In such multilayers, excitation with a single ultrashort laser pulse creates a skyrmion state \cite{Je2018, Buttner2020} with a density controlled by the applied magnetic field \cite{Gerlinger2021}.
We confirmed this for our sample via static x-ray microscopy measurements (Fig.~\ref{fig:expdata}(c,d), SM section I, and Ref.~\citenum{Kern2022_2}).

To detect the transient fluctuation state, we performed time-resolved measurements by probing lateral spin textures with resonant small-angle x-ray scattering (SAXS) in transmission after IR excitation at \SI{795}{\nano\meter} wavelength with \SI{65}{\femto\second} (fwhm) pulses (Fig.~\ref{fig:expdata}(a)). Importantly, we varied the applied fields during the measurements (indicated as blue dots in Fig.~\ref{fig:expdata}(b)), corresponding to different skyrmion densities in the final states down to saturated states.
Experiments were conducted at the DiProI beamline of the free-electron laser (FEL) source FERMI@Elettra, providing sub-\SI{100}{\femto\second} (fwhm) pulses of XUV radiation \cite{Capotondi2013}.
To establish resonant scattering conditions, the FEL was tuned to the Co M$_{2,3}$ edge (\SI{20.8}{nm} wavelength), resulting in a sensitivity to out-of-plane magnetic moments (see SM section I for details).
As a typical experimental procedure, we saturated the sample at \SI{\approx 220}{\milli\tesla} in between two FEL shots and set the desired external field for the subsequent pump--probe measurement in an automated field cycling synchronized to the FEL repetition rate of \SI{50}{\hertz} (see supplemental Fig.~S1).


Azimuthal integrations of typical scattering patterns recorded at different delay times are shown in Fig.~\ref{fig:expdata}(e,f) for \SI{34}{\milli\tesla} and \SI{197}{\milli\tesla}, respectively, as a function of the momentum transfer $q$ (see supplemental Fig.~S4 for all integrations).
The magnetic film and the 30-nm thick silicon-nitride substrate are laterally almost homogeneous and scattering only originates from lateral variations of the magnetization.

For both fields shown in Fig.~\ref{fig:expdata}(e,f), we observe that \SI{100}{\pico\second} after IR excitation, the scattering spans across a large range in momentum space, in particular indicating the presence of short-range spin correlations during the fluctuation state \cite{Buttner2020}.
At these early delay times, we observe that the shape of the SAXS spectra is almost independent of the applied magnetic field, with similar total scattering intensity.
At later times, however, the evolution of the scattering patterns is very different depending on the applied field.
At low field, the scattering intensity distribution narrows and shifts its maximum towards lower scattering vectors.
At the same time, the intensity rapidly increases, indicating the progressing formation and growth of skyrmions.
The very prominent scattering-intensity maximum of the resulting final state at delays \SI{>1}{\nano\second} reveals the creation of a skyrmion pattern with a high degree of long-range ordering in agreement with the corresponding real-space image (Fig.~\ref{fig:expdata}(c)).
In contrast, for a high applied field, the sample ultimately returns to the uniform saturated state, which is evidenced by a complete loss of scattering intensity.

\begin{figure}
    \includegraphics[width=\columnwidth]{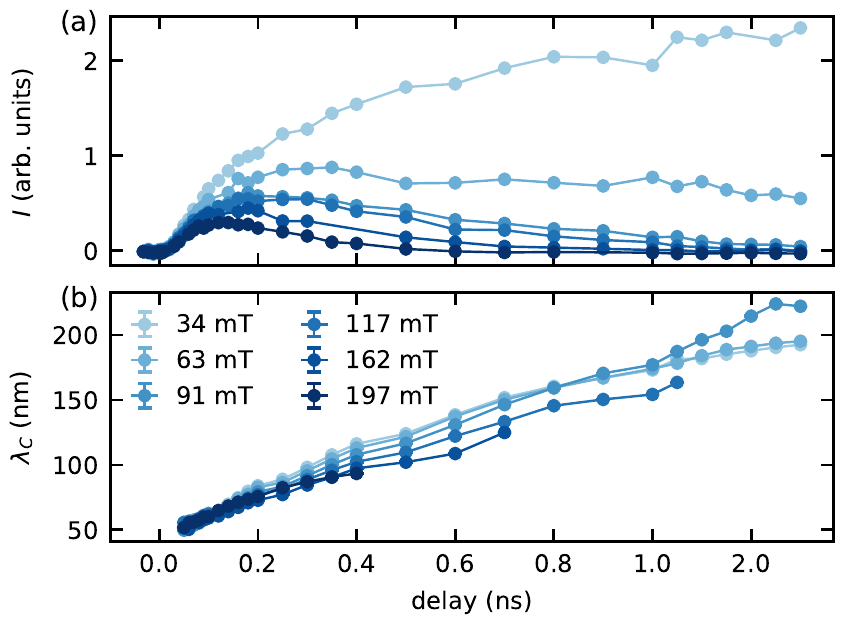}
    \caption{Results from the pump--probe scattering experiments. (a) The total scattered intensity $I$ and (b) the correlation length ($\lambda_C = 2\pi/q_0$), for six different applied magnetic fields as a function of the delay between IR pump and FEL probe.}
    \label{fig:expresults}
\end{figure}

The entire picosecond evolution of the total scattering intensity is shown in Fig.~\ref{fig:expresults}(a) as a function of the pump--probe delay for different applied fields.
Notably, in the first tens of picoseconds after excitation, the scattered intensity increases with a rate that is almost independent of the applied field.
At lower fields, promoting a stable final skyrmion state, the intensity keeps growing also on time scales of several hundreds of picoseconds.
On the other hand, the scattering intensity again completely declines after a maximum at around \SI{150}{\pico\second} for higher fields where the film finally transitions back to the saturated state.
The large variation of scattering intensity in the final states directly reflects the field dependence of the skyrmion density \cite{Gerlinger2021}.

In the next step, we deduce an estimate for the dominant length scales of the lateral spin correlations causing the emerging magnetic scattering intensity.
A common measure for this length scale is the correlation length---the inverse $q$-space position of the SAXS intensity maximum ($q_0$), plotted in Fig.~\ref{fig:expresults}(b).
Once an intensity maximum has formed after \SI{50}{\pico\second}, the correlation length steadily increases with time.
It is remarkable that this evolution is almost entirely independent of the applied field, in particular at early times (\SI{<300}{\pico\second}).

Despite the high similarity of all SAXS spectra at early times (\SI{<300}{\pico\second}), the SAXS spectra at late times exhibit significant differences at higher momentum transfer---a region where the form factor of the skyrmions dominates the scattering (Fig.~\ref{fig:expdata}(e,f)).
We fitted the average form factor assuming circular textures with normally distributed sizes to the SAXS spectra of the three final states containing skyrmions (SM, section II, Fig.~S6). 
The radius decreases with the applied magnetic field as is expected from theory \cite{Buttner2018} and previous measurements on similar samples \cite{Gerlinger2021, Kern2022_1}.

It is already known that the early times of the skyrmion formation process are governed by spin and topological fluctuations while later times are characterized by a relaxation of the state to the final skyrmion size and density \cite{Buttner2020}.
We, here, again clearly detect this fluctuation state based on the evolution of the scattering intensity and in-plane correlation length.
The behavior at early times (\SI{<300}{\pico\second}) shows that the magnetic film transiently enters the topological fluctuation state no matter if a stable skyrmion state is finally formed or not.
The final formation of skyrmions is, thus, not a necessary condition for the transient occurrence of the fluctuation state.
Under suitable conditions, transient topological fluctuations do exist even if ultimately the topology remains unchanged. 
Only at later times (\SI{>300}{\pico\second}), the applied field governs the relaxation process and the resulting final state.


\begin{figure}
    \includegraphics[width=\columnwidth]{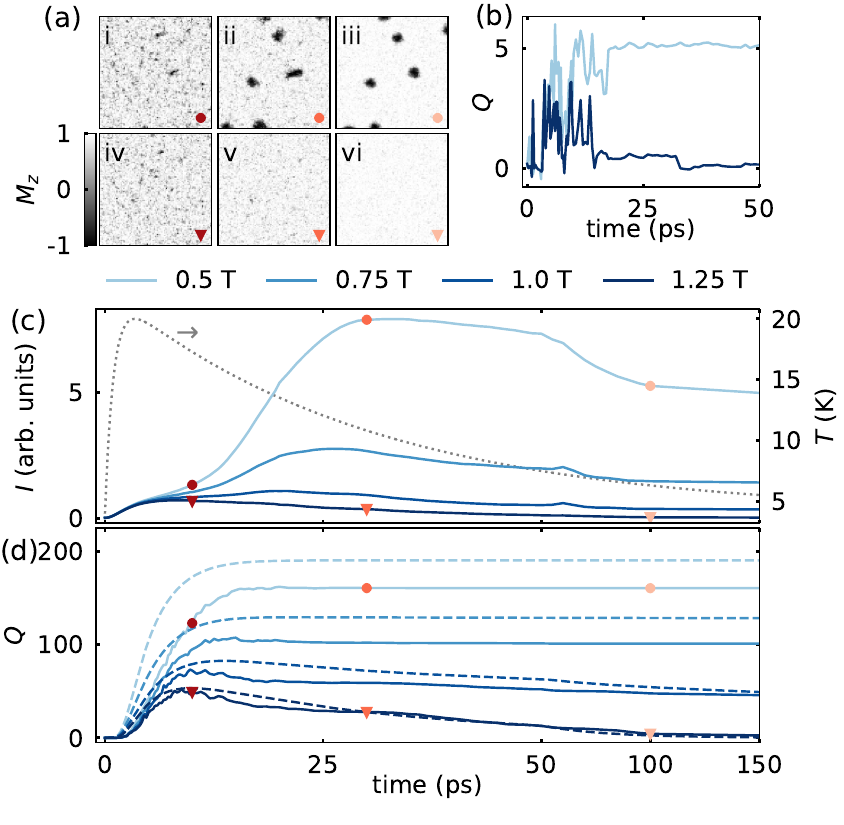}
    \caption{Results from the atomistic spin simulations. (a) \SIrange[range-phrase=$\times$, range-units=single]{100}{100}{spins} snapshots of $M_{iz}$ at \SIlist{10;30;100}{\pico\second} as indicated in (c) and (d) at \SI{0.5}{\tesla} (i--iii, circles) and \SI{1.25}{\tesla} (iv--vi, triangles).
    (b) Time evolution of the topological charge $Q$ in the images shown in (a).
    Time evolution of (c) the intensity $I$ of the Fourier transform of $M_{iz}$ and (d) the topological charge $Q$, mean from \num{10} simulation runs of \SIrange[range-phrase=$\times$, range-units=single]{500}{500}{spins} at different magnetic fields.
    The temperature pulse is shown as a dotted gray line in (c).
    The dashed lines in (d) show the predictions of the topological charge obtained from a rate-equation model.
    }
    \label{fig:simIQ}
\end{figure}

We will interpret these results using atomistic simulations, assuming a minimal model for supporting skyrmions with only short-range interactions, which was previously found to be sufficient to qualitatively understand the short-time nucleation dynamics \cite{Buttner2020} (more details in SM section III).
We performed simulations at varying magnetic fields with a temperature pulse to mimic the laser excitation (shown as a gray dotted line in Fig.~\ref{fig:simIQ}(c)). 
Example snapshots from the simulation of the spin's z-component ($M_{iz}$) with progressing simulation time are shown in Fig.~\ref{fig:simIQ}(a) for the lowest and highest magnetic field used.
In agreement with the experiment, the system develops skyrmions at the lowest field, while at high applied field, the system returns to a saturated state.

The azimuthally integrated spectra of the intensity of the Fourier transform $\mathcal{F}$ of $M_{iz}$ (shown in the supplemental Fig.~S7) behave similarly to the experimental data (Fig.~\ref{fig:expdata}(f,g)) but are not identical in shape due to the lack of long-range interactions.
Fig.~\ref{fig:simIQ}(c) shows the sum of $|\mathcal{F} (M_{iz})|^2$ (called intensity) which evolves very similarly to the totally scattered intensity of our experiment (Fig.~\ref{fig:expresults}(a)).
In particular, the intensity increase is independent of the magnetic field in the first few picoseconds and is later directly related to the density of skyrmions nucleated in the final state.

In addition to providing a real-space picture of the magnetization textures formed, the simulations also give access to the topological charge $Q$ of the system.
The evolution of $Q$ averaged over ten simulation runs is shown as solid lines in Fig.~\ref{fig:simIQ}(d).
Already before the maximum temperature is reached, topological charge is created and starts to fluctuate at all fields.
The period the system stays in the fluctuation state can be clearly identified by rapid changes in the topological charge as exemplarily shown in Fig.~\ref{fig:simIQ}(b).
For low magnetic fields, the topological charge stays constant after the fluctuation state whereas, at high fields, $Q$ declines as does the intensity $I$, both displaying the disappearance of topological spin textures.
This finding is again in line with our experimental observations.

To obtain a better understanding of the field dependence of the fluctuation state, we performed another set of atomistic simulations where we let the system relax to equilibrium at constant temperature either from a spin-aligned ferromagnetic (FM) phase or from a random paramagnetic (PM) phase (SM section III).
We reproduce previous findings of an intermediate (IM) phase between the FM and the PM phases at temperatures close to the Curie temperature $T_\text{C}$ featuring skyrmions with a finite lifetime \cite{Boettcher2018, Buttner2020}.
Furthermore, we find that the topological charge $Q$ can be described sufficiently accurately by the first-order approximation for the relaxation 
\begin{equation}
    \frac{\mathrm{d}Q}{\mathrm{d}t} = \frac{Q_\text{f} - Q}{\tau_\text{C}},
    \label{eq:Qt}
\end{equation}
where $Q_\text{f}$ denotes the topological charge at equilibrium and $\tau_\text{C}$ the relaxation timescale.
Both $Q_\text{f}$ and $\tau_\text{C}$ are dependent on the magnetic field $B$ and the temperature $T$ and can be obtained from fits of the solution of Eq.~\eqref{eq:Qt} to the relaxation data (see supplemental Fig.~S8).
Using temperature-interpolated data for $Q_\text{f}$ and $\tau_\text{C}$, the basic rate equation in Eq.~\eqref{eq:Qt} can be solved even for a (time-varying) temperature pulse to predict the topological charge after the pulse at a given magnetic field.
These predictions, shown in Fig.~\ref{fig:simIQ}(c) as dashed lines, qualitatively model the data extracted from the atomistic simulations with a temperature pulse.

Looking closer at the relaxation time scale $\tau_\text{C}$, we find that it diverges at low temperatures and low magnetic fields.
At high temperatures, i.e., in the IM phase, the relaxation time scales are of the order of picoseconds for all fields, leading to a universal emergence of topological charge at the peak of the temperature pulse.
When the temperature drops and the system is kept under low magnetic fields, the system cannot relax to the equilibrium FM state and skyrmions effectively freeze out of the fluctuation state due to the large relaxation time scales.
At high fields, however, $\tau_\text{C}$ is short enough for the system to relax to the FM state.
The topological switching is, therefore, a non-equilibrium phase transition that happens on time scales that are much shorter than the relaxation time of the system.

\begin{figure*}
    \includegraphics[width=2\columnwidth]{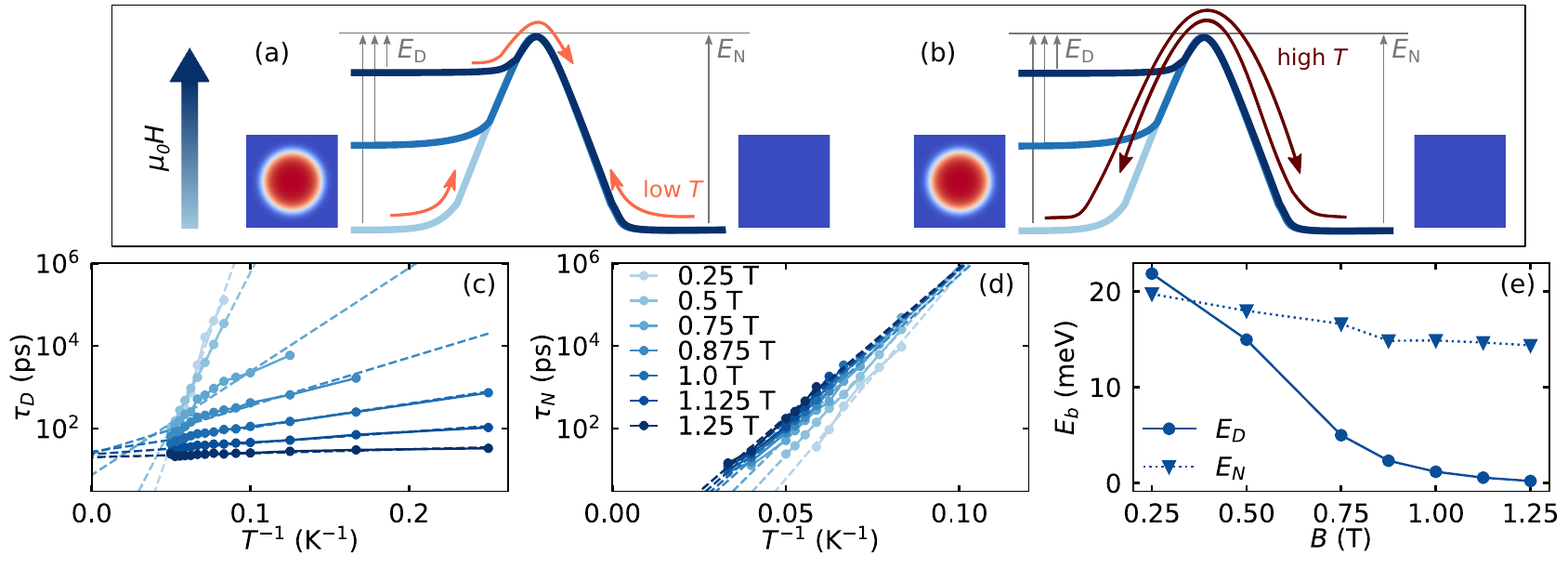}
    \caption{Energy barriers for skyrmion nucleation and decay extracted from Arrhenius model. 
    (a,b) Schematic of the field-dependent energy barriers (blue curves): (a) at low temperatures skyrmion decay is only possible at high fields; (b) at high temperatures skyrmions nucleate and decay at all fields because the thermal fluctuations (red arrows) are large.
    (c), (d) Decay and nucleation time scales $\tau_\text{D}$ and $\tau_\text{N}$, respectively, as a function of the inverse temperature.
    Dashed lines represent Arrhenius fits.
    (e) The energy barriers for the decay (circles, solid line) and the nucleation (triangles, dotted line) determined from the Arrhenius fit as a function of the magnetic field.}
    \label{fig:Eb}
\end{figure*}

As the fluctuation state is characterized by rapid nucleation and decay of topological charge \cite{Buttner2020}, we now divide the relaxation time $\tau_\text{C}$ into two independent time scales: a nucleation time $\tau_\text{N}$ and a decay time $\tau_\text{D}$ related to $\tau_\text{C}$ as $1/\tau_\text{C} = 1/\tau_\text{N} + 1/\tau_\text{D}$.
Since the balance of nucleation and decay determines the equilibrium topological charge, $Q_\text{f}$ is also a function of $\tau_\text{N}$ and $\tau_\text{D}$, which allows us to find $\tau_\text{N}$ and $\tau_\text{D}$ from $Q_\text{f}$ and $\tau_\text{C}$ as determined from the simulations (see SM, section III, for complete derivation).
We show the results for $\tau_\text{D}$ and $\tau_\text{N}$ in Fig.~\ref{fig:Eb}(c) and (d), respectively, and observe that both relaxation times are well described by an Arrhenius law \cite{Rozsa2016, Wild2017, Bessarab2018}.
The resulting field-dependent energy barriers are shown in Fig.~\ref{fig:Eb}(e).
The energy barrier for the skyrmion nucleation $E_\text{N}$ is almost independent of the magnetic field, which explains why the fluctuation state can robustly be accessed via photoexcitation independently of the applied field as expected \cite{Buttner2018}. 
However, the energy barrier for the skyrmion decay $E_\text{D}$ declines with the field and is almost zero for the highest magnetic fields considered here.
This explains why, 
at high magnetic fields,
the skyrmions do not freeze out like at lower fields.
We illustrate the field-dependent barriers and corresponding fluctuations at low and high temperature in Fig.~\ref{fig:Eb}(a,b). 

Interestingly, the energy barriers extracted from the relaxation simulations have the same order of magnitude and show the same dependence on the magnetic field as predictions for the energy barrier at zero temperature (SM, section III) \cite{Oik2016, Schuette2014, Bessarab2018, Buttner2018, Belavin1975, Tretiakov2007}.
This seemingly disproves the previous suggestion that the fluctuation state is primarily initiated by a substantially reduced energy barrier compared to the ground state \cite{Buttner2020}. 
In fact, it is surprising that such high energy barriers are present in the high-temperature, fluctuation-disordered phase \cite{Rozsa2016, Boettcher2018}, where skyrmions have a short lifetime. 
However, we note that the pre-exponential factor (attempt frequency) is much higher than the commonly accepted micromagnetic estimate of \SI{\approx 1e9}{Hz} (see supplemental Fig.~S11). 
Such substantially increased values for the attempt frequency, including a strong dependence on the applied field over orders of magnitude, were also found in previous calculations \cite{Bessarab2018} and experiments \cite{Wild2017} investigating skyrmion stability and have been traced back to entropic effects, which indeed differentiate between continuum and lattice models.
Interestingly, our results evidence that physical mechanisms elucidated before at low temperatures (far from the Curie temperature) and on long timescales (down to milliseconds) can also dictate topological fluctuation dynamics at high temperature and on ultrafast timescales.

Nevertheless, we also emphasize that our atomistic simulations only feature a simplified, minimal model for the skyrmion nucleation dynamics.
In the multilayer samples used in the experiments, the energy barrier is additionally determined by the alignment between the ferromagnetic layers, which is absent in our single-layer model.
It is likely that eliminating the layer coupling will lead to a reduction of the energy barriers at high temperatures, promoting topological switching \cite{Buttner2018, Buttner2020}.
While this effect would additionally contribute to the high nucleation and decay rates in the fluctuation states, it is insufficient to explain the freeze out of skyrmions when the temperature drops again.
In the ferromagnetic multilayer, the skyrmion state is stabilized by stray fields even in the ground state.
This long-range interaction is missing in the atomistic simulations, and the frozen skyrmion state is considered metastable.
However, our rate model successfully explains the experimentally confirmed, transient formation of skyrmions at applied fields supporting a saturated ground state.
Moreover, being based on only minimal assumptions, the model suggests similar ultrafast laser-induced topological switching and skyrmion formation mechanisms also for ferrimagnetic \cite{Caretta2018} and antiferromagnetic \cite{Juge2022} films with a small or vanishing net magnetization.

In conclusion, we here find that the fluctuation state 
can robustly be accessed by ultrashort laser excitation.
While the fluctuation state promotes the topological switching into a skyrmions state, it also appears independently of this transition (at high applied fields).
We attribute this robustness of the fluctuation state to a field-independent nucleation energy barrier, whereas the topology of the final state is determined by diverging relaxation time scales,  resulting from a field-dependent skyrmion decay energy barrier (Fig.~\ref{fig:Eb}(a,b)).
Given the minimal assumptions in our model, the reported asymmetric energy barrier for nucleation and decay indicates a general mechanism for the creation of stable topological textures from non-equilibrium states in other magnetic materials as discussed.
This potentially includes even systems with entirely different competing interactions, like polar materials \cite{Das2019}.
Further research on the fluctuation state will have to address the role of long-range stray fields and interlayer coupling, which ultimately may facilitate the discovery of the fundamental speed limit of the topological transition in ferromagnetic multilayers and related materials.

\section*{Acknowledgements}
We acknowledge FERMI@Elettra Trieste for providing access to its free electron laser facility.
The authors thank Ingo Will from the Max Born Institute in Berlin and Sebastian Wintz from the Max Planck Institute for Intelligent Systems in Stuttgart for support with the measurements at MAXYMUS.
This research is part of the Shell NWO/FOM inititative ``Computational sciences for energy research'' of Shell and Chemical Sciences, Earth and Life Sciences, Physical Sciences, Stichting voor Fundamenteel Onderzoek der Materie (FOM) and Stichting voor de Technische Wetenschappen (STW).
Financial support from the Leibniz Association via Grant No. K162/2018 (OptiSPIN), the Helmholtz Young Investigator Group Program via Grant No. VH-NG-1520, and the European Research Council under ERC Grant Agreement No. 856538 (3D-MAGiC) is acknowledged.

\section*{Data Availability Statement}

The data of this study is available from the corresponding author upon reasonable request.

\bibliography{literature}

\end{document}